\documentstyle[aps]{revtex}

\begin{document}
\title{Collapsing strange quark matter in Vaidya Geometry}
\author{T. Harko\footnote{E-mail: tcharko@hkusua.hku.hk}}
\address{Department of Physics, The University of Hong Kong,
         Pokfulam, Hong Kong}
\author{K. S. Cheng\footnote{E-mail: hrspksc@hkucc.hku.hk}}
\date{24 April 2001}
\maketitle

\begin{abstract}
Exact solutions of the gravitational field equations for a mixture of a null
charged strange quark fluid and radiation are obtained in a Vaidya space-time.
The conditions for the formation of a naked singularity are analyzed by considering
the behavior of radial geodesics originating from the central singularity.
\end{abstract}

\section{Introduction}

It is generally believed that strange quark matter, consisting of $%
u,d$ and $s$ quarks, is the most energetically favorable state of baryon
matter. Witten [1] suggested that there are two ways of formation of the
strange matter: the quark-hadron phase transition in the early universe and
the conversion of neutron stars into strange ones at ultrahigh densities.
Some theories of strong interactions, e.g. quark bag models, suppose that
the breaking of physical vacuum takes place inside hadrons. As a result the
vacuum energy densities inside and outside a hadron become essentially
different and the vacuum pressure on a bag wall equilibrates the pressure of
quarks thus stabilizing the system.

There are several proposed mechanisms for the formation of quark stars.
Quark stars are expected to form during the collapse of the core of a
massive star after the supernova explosion as a result of a first or second
order phase transition, resulting in deconfined quark matter [2]. The
proto-neutron star core or the neutron star core is a favorable environment
for the conversion of ordinary matter to strange quark matter [3]. Another
possibility for strange star formation is that some neutron stars in
low-mass X-ray binaries can accrete sufficient mass to undergo a phase
transition to become strange star [4]. Such a possibility is also supported
by an unusual hard X-ray burster [5]. This mechanism has been proposed as a
source of radiation emission for cosmological $\gamma $-ray bursts [6].
Other possible astrophysical phenomena related to strange stars are
discusssed in Ref. [2]. Hence the problem of the collapse of the strange
matter is of much interest in the study of strange quark star formation,
dynamics and evolution. On the other hand the theoretical understanding of
the collapse is of fundamental importance in general relativity.

Usually it is considered that quark matter is formed from a Fermi gas of $3A$
quarks constituting a single color-singlet baryon with baryon number $A$.
The theory of the equation of state of strange matter is directly based on
the fundamental QCD Lagrangian [7]
\begin{equation}
L_{QCD}=\frac{1}{4}\sum_{a}F_{\mu \nu }^{a}F^{a\mu \nu }+\sum_{f=1}^{N_{f}}%
\bar{\psi}\left( i\gamma ^{\mu }\partial _{\mu }-g\gamma ^{\mu }A_{\mu }^{a}%
\frac{\lambda ^{a}}{2}-m_{f}\right) \psi .
\end{equation}

The subscript $f$ denotes the various quark flavors $u,d,s$ etc. and the
nonlinear gluon field strength is given by [7]
\begin{equation}
F_{\mu \nu }^{a}=\partial _{\mu }A_{\nu }^{a}-\partial _{\nu }A_{\mu
}^{a}+gf_{abc}A_{\mu }^{b}A_{\nu }^{c}.
\end{equation}

QCD predicts a weakening of the quark-quark interaction at short distances
(or high momenta $Q^{2}$), because the one-loop series for the gluon
propagator yields a running coupling constant.

Neglecting quark masses in first order perturbation theory, the equation of
state for zero temperature quark matter is [1,7]
\begin{equation}\label{1}
p=\frac{1}{3}\left( \rho -4B\right) ,  
\end{equation}
where $B$ is the difference between the energy density of the perturbative
and non-perturbative QCD vacuum (the bag constant) and $\rho ,p$ are the
energy density and thermodynamic pressure of the quark matter, respectively.
Eq. (\ref{1}) is essentially the equation of state of a gas of massless particles
with corrections due to the QCD trace anomaly and perturbative interactions.
The vacuum pressure $B$, which holds quark matter together, is a simple
model for the long-range, confining interactions in QCD. At the surface of
the star, as $p\rightarrow 0$, we have $\rho \rightarrow 4B$. The typical
value of the bag constant is of the order $B=57MeV/fm^{3}\approx
10^{15}g/cm^{3}$ [2]. After the neutron matter-quark matter phase transition
(which is supposed to take place in the dense core of neutron stars) the
energy density of strange matter is $\rho \approx 5\times 10^{14}g/cm^{3}$
[1].Therefore quark matter always satisfy the condition $p\geq 0$.

It is the purpose of the present Letter to obtain an exact non-static
solution of the Einstein field equations for a collapsing charged null
strange fluid in a Vaidya space-time and to study some of its singularity
properties. The Vaidya geometry permitting the incorporations of the effects
of radiation offers a more realistic background than static geometries,
where all back reaction is ignored. The exact solution obtained represents
the generalization, for strange matter, of the collapsing solutions
previously obtained by Vaidya [8], Bonnor and Vaidya [9], Lake and Zannias
[10] and Husain [11]. For a general method of obtaining spherically
symmetric solutions in Vaidya geometry, see Ref. [12]. The structure and
properties of singularities in the gravitational collapse in Vaidya
space-times have been analyzed in Refs. [13-17].

The present Letter is organized as follows. In Section II the gravitational
field equations are written down and the general solution is obtained. The
conditions of formation of a naked singularity are analyzed in Section III. In
Section IV we discuss and conclude our results.

\section{Spherical collapse of the strange quark null fluid}

In ingoing Bondi coordinates ($u,r,\theta ,\varphi $) and with advanced
Eddington time coordinate $u=t+r$ (with $r\geq 0$ the radial coordinate and $%
r$ decreasing towards the future) the line element describing the radial
collapse of a coherent stream of charged strange matter is
\begin{equation}
ds^{2}=-\left[ 1-\frac{2m\left( u,r\right) }{r}\right] du^{2}+2dudr+r^{2}%
\left( d\theta ^{2}+\sin ^{2}\theta d\varphi ^{2}\right) .  \label{2}
\end{equation}

$m\left( u,r\right) $ is the mass function and gives the gravitational mass
within a given radius $r$.

Equilibrium configuration of masless strange quark matter has equal numbers
of $u,d$ and $s$ quarks and is electrically neutral. If the mass of the
strange quark is not zero, strange quarks are depleted and the system
develops a net positive charge. Since stars in their lowest energy state are
supposed to be charge neutral, electrons must balance the net positive quark
charge in strange matter and the condition of charge neutrality requires $%
\frac{2}{3}n_{u}-\frac{1}{3}n_{d}-\frac{1}{3}n_{s}-n_{e}=0$, \ where $n$ is
the particle number density [2]. Hence quark matter is usually neutral. In
principle, rather extreme astrophysical conditions could lead to a charged
strange quark matter astrophysical configuration and, for the sake of
generality, we shall also consider this possibility.

We shall write the matter energy-momentum tensor in the form [11,12]
\begin{equation}\label{3}
T_{\mu \nu }=T_{\mu \nu }^{(n)}+T_{\mu \nu }^{(m)}+E_{\mu \nu },
\end{equation}
where
\begin{equation}
T_{\mu \nu }^{(n)}=\mu \left( u,r\right) l_{\mu }l_{\nu },
\end{equation}
is the component of the matter field that moves along the null hypersurfaces 
$u=const.$,
\begin{equation}
T_{\mu \nu }^{(m)}=\left( \rho +p\right) \left( l_{\mu }n_{\nu }+l_{\nu
}n_{\mu }\right) +pg_{\mu \nu },  \label{5}
\end{equation}
represents the energy-momentum tensor of the strange quark matter and

\begin{equation}
E_{\mu \nu }=\frac{1}{4\pi }\left( F_{\mu \alpha }F_{\nu }^{\alpha }-\frac{1%
}{4}g_{\mu \nu }F_{\alpha \beta }F^{\alpha \beta }\right) , 
\end{equation}
is the electromagnetic contribution. $l_{\mu }$ and $n_{\mu }$ are two
null-vectors given by $l_{\mu }=\delta _{\left( \mu \right) }^{\left( \mu
\right) }$ and $n_{\mu }=\frac{1}{2}\left[ 1-\frac{2m\left( u,r\right) }{r}%
\right] \delta _{\left( \mu \right) }^{\left( \mu \right) }-\delta _{\left(
\mu \right) }^{\left( 1\right) }$, so that $l_{\alpha }l^{\alpha }=n_{\alpha
}n^{\alpha }=0$ and $l_{\alpha }n^{\alpha }=-1$ [11,12] (with $\delta
_{\left( b\right) }^{\left( a\right) }$ the Kronecker symbol). The energy
density and pressure in Eq. (\ref{5}) have been obtained by diagonalizing the energy-momentum tensor
obtained from the metric [11]. The electromagnetic tensor $F_{\mu \nu }$
obeys the Maxwell equations
\begin{equation}
\frac{\partial F_{\mu \nu }}{\partial x^{\lambda }}+\frac{\partial
F_{\lambda \mu }}{\partial x^{\nu }}+\frac{\partial F_{\nu \lambda }}{%
\partial x^{\mu }}=0,  \label{7}
\end{equation}
\begin{equation}
\frac{1}{\sqrt{-g}}\frac{\partial }{\partial x^{\mu }}\left( \sqrt{-g}F^{\mu
\nu }\right) =-4\pi j^{\mu }.  \label{8}
\end{equation}

Without any loss of generality we take the vector potential to be [10]
\begin{equation}
A_{\mu }=\frac{q\left( u\right) }{r}\delta _{\left( \mu \right) }^{\left(
\mu \right) },
\end{equation}
with $q(u)$ being an arbitrary function. From the Maxwell Eqs. (\ref{7}), (\ref{8}) it follows that the only non-vanishing components of $F_{\mu \nu }$
are $F_{ru}=-F_{ur}=\frac{q(u)}{r^{2}}$ and
\begin{equation}
E_{\mu }^{\nu }=\frac{q^{2}(u)}{r^{4}}diag\left( -1,1,-1,1\right) .
\end{equation}

For the energy-momentum tensor (\ref{3}) the gravitational field equations take
the form
\begin{equation}
\frac{1}{r^{2}}\frac{\partial m\left( u,r\right) }{\partial u}=4\pi \mu
\left( u,r\right) ,
\end{equation}
\begin{equation}
\frac{2}{r^{2}}\frac{\partial m(u,r)}{\partial r}=8\pi \rho \left(
u,r\right) +\frac{q^{2}(u)}{r^{4}},
\end{equation}
\begin{equation}
-\frac{1}{r}\frac{\partial ^{2}m\left( u,r\right) }{\partial r^{2}}=8\pi
p\left( u,r\right) +\frac{q^{2}(u)}{r^{4}}.
\end{equation}

By using the bag equation of state of strange matter given by Eq. (\ref{1}), we obtain the following equation describing the dynamics of a
mixture of null fluid, strange matter and electric field:
\begin{equation}
\frac{\partial ^{2}m\left( u,r\right) }{\partial r^{2}}=-\frac{2}{3r}\frac{%
\partial m(u,r)}{\partial r}-\frac{2}{3}\frac{q^{2}(u)}{r^{3}}+\frac{32\pi B%
}{3}r.
\end{equation}

By means of the substitution
\begin{equation}
m\left( u,r\right) =m_{0}\left( u,r\right) +\frac{4\pi B}{3}r^{3}-\frac{%
q^{2}(u)}{2r},
\end{equation}
we obtain a new unknown function $m_{0}\left( u,r\right) $, which satisfies
the equation
\begin{equation}
\frac{\partial ^{2}m_{0}\left( u,r\right) }{\partial r^{2}}=-\frac{2}{3r}%
\frac{\partial m_{0}(u,r)}{\partial r},
\end{equation}
having the general solution
\begin{equation}
m_{0}\left( u,r\right) =C\left( u\right) r^{1/3}+F(u),
\end{equation}
with $C\left( u\right) $ and $F(u)$ being two arbitrary functions. Hence the
general solution of the gravitational field equations for collapsing
strange matter in the Vaidya metric (\ref{2}) is given by
\begin{equation}
m\left( u,r\right) =F(u)+C\left( u\right) r^{1/3}+\frac{4\pi B}{3}r^{3}-%
\frac{q^{2}(u)}{2r},  \label{18}
\end{equation}
\begin{equation}
\mu \left( u,r\right) =\frac{1}{4\pi r^{2}}\left[ \frac{dF(u)}{du}+\frac{%
dC(u)}{du}r^{1/3}-q(u)\frac{dq(u)}{du}\frac{1}{r}\right] ,
\end{equation}
\begin{equation}
\rho \left( u,r\right) =\frac{1}{4\pi r^{2}}\left[ \frac{1}{3}C(u)r^{-2/3}+%
\frac{q^{2}(u)}{2r^{2}}+4\pi Br^{2}\right] ,
\end{equation}
\begin{equation}
p\left( u,r\right) =\frac{1}{12\pi r^{2}}\left[ \frac{1}{3}C(u)r^{-2/3}+%
\frac{q^{2}(u)}{2r^{2}}-12\pi Br^{2}\right] .  \label{21}
\end{equation}

The electromagnetic current follows from the Maxwell Eq. (\ref{8}) and is given by
\begin{equation}
j^{\mu }=\frac{1}{4\pi r^{2}}\frac{dq(u)}{du}l^{\mu }.
\end{equation}

For $B=0$ and $q(u)=0$ we obtain the solution given by Husain [11] for the
null fluid pressure satisfying $p=\frac{1}{3}\rho $.

The energy-momentum tensor of the mixture of fluids under consideration
belongs to the Type II fluids [11]. The energy conditions are the weak,
strong and dominant energy conditions $\mu \geq 0,\rho \geq 0,p\geq 0,\rho
\geq p\geq 0$ and can be satisfied by appropriately choosing the arbitrary
functions $F(u)$ and $C(u)$ that characterize the injection and initial
distribution of mass and $q(u)$ that describes the variation of the charge.
The condition $\mu \geq 0$ is equivalent to $\frac{dm}{du}\geq 0$ and leads
to
\begin{equation}\label{22}
\frac{dF(u)}{du}+\frac{dC(u)}{du}r^{1/3}\geq q(u)\frac{dq(u)}{du}\frac{1}{r},
\end{equation}
imposing a simultaneous constraint on all three functions $F(u),C(u)$ and $%
q(u)$. For small values of $r$ and for charged quark matter, the right hand
side of Eq. (\ref{22}) dominates and this energy condition could not hold. One
possibility to satisfy Eq. (\ref{22}) for all $r$ is to assume that the function $q(u)$ behaves so that 
$\frac{dq^{2}(u)}{du}\rightarrow 0$ for $r\rightarrow 0$. This means that
the charge in the singular point $r=0$ is constant for all times.
Alternatively, we may suppose that at extremely small radii matter is
converted to strange quark matter so as to satisfy the energy condition. For
neutral quark matter Eq. (\ref{22}) is easily satisfyed by choosing $\frac{dF(u)}{du}>0$ and $\frac{%
dC(u)}{du}>0$. To satisfy the condition $p\geq 0$ for large $r$, we must
impose on the function $C(u)$ the constraint $C(u)\geq 36\pi Br^{7/2}\geq
0,\forall u$. With this choice the condition of the non-negative energy
density is also automatically satisfied. Due to the bag equation of state (\ref{1}) we always have $\rho \geq p\geq 0$ and thus the dominant energy
condition holds, too.

The radii of the apparent horizon of the metric (\ref{2}) are given by the solution of the equation $2m=r$. If $%
\lim_{u\rightarrow \infty }F(u)=F_{0}=const.$, $\lim_{u\rightarrow \infty
}C(u)=C_{0}=const.$ and $\lim_{u\rightarrow \infty }q(u)=q_{0}=const.$ then
the algebraic equation determining the radii of the apparent horizons is
\begin{equation}
2F_{0}+2C_{0}r^{1/3}+\frac{8\pi B}{3}r^{3}-\frac{q_{0}^{2}}{r}=r,
\end{equation}
which in general may have multiple solutions.

The singularities of the quark matter filled Vaidya space-time can be
recognized from the behavior of the energy density and curvature scalars
like e.g. $R_{\alpha \beta }R^{\alpha \beta }$ given by
\begin{equation}
R_{\alpha \beta }R^{\alpha \beta }=\frac{8}{r^{4}}\left[ \frac{1}{3}%
C(u)r^{-8/3}+\frac{q^{2}(u)}{r^{4}}+4\pi B\right] ^{2},
\end{equation}
which diverges for $r\rightarrow 0$.

\section{Outgoing radial null geodesics equation}

The central shell-focusing singularity (i.e. that occuring at $r=0$) is
naked if the radial null-geodesic equation admits one or more positive real
root $X_{0}$ [16]. In the case of the pure Vaidya space-time it has been
shown that for a linear mass function $2m(u)=\lambda u$ the singularity at $%
r=0,u=0$ is naked for $\lambda \leq \frac{1}{8}$ [17]. In order to simplify
calculations we choose some particular expressions for the functions $%
F(u),C(u)$ and $q(u)$, e.g. $F(u)=\alpha u/2$, $C(u)=\beta u^{2/3}$ and $%
q(u)=\gamma u$, \ with $\alpha >0,\beta >0$ and $\gamma \geq 0$ constants.
With this choice the equation of the radially outgoing, future-directed null
geodesic originating at the singularity is
\begin{equation}
\frac{du}{dr}=\frac{2}{1-\frac{\alpha u}{r}-\beta \left( \frac{u}{r}\right)
^{2/3}-\gamma \left( \frac{u}{r}\right) ^{2}-\frac{8\pi B}{3}r^{2}}.
\label{24}
\end{equation}

For the geodesic tangent to be uniquely defined and to exist at the singular
point $r=0,u=0$ of Eq. (\ref{24}) the following condition must hold [16]:
\begin{equation}\label{25}
\lim_{u,r\rightarrow 0}\frac{u}{r}=\lim_{u,r\rightarrow 0}\frac{du}{dr}%
=X_{0}.  
\end{equation}

When the limit exists and $X_{0}$ is real and positive, there is a future
directed, non-space=like geodesic originating from $r=0,u=0$. In this case
the singularity will be, at least, locally naked. For the null geodesic Eq. (%
\ref{24}) condition (\ref{25}) leads to the follosing algebraic equation:
\begin{equation}
\gamma X_{0}^{3}+\alpha X_{0}^{2}+\beta X_{0}^{5/3}-X_{0}+2=0.  \label{26}
\end{equation}

With the help of the substitution $X_{0}=y^{3}$ Eq. (\ref{26}) becomes a ninth order polynomial equation of the form $%
f(y)=\gamma y^{9}+\alpha y^{6}+\beta y^{5}-y^{3}+2=0$. According to a
theorem given by Poincare [18] the number of positive roots of a polynomial
equation equals the number of changes in sign in the sequence of the
non-negative coefficients of the polynomial $g(y)=\left( 1+y\right) ^{k}f(y)$%
. The Poincare criterion indicates that Eq. (\ref{26}) has two positive roots. In order to give an estimate of the
positive roots we shall consider the expression $g(y)=\left( 1+y\right)
^{k}f(y)=\sum_{\nu =0}^{k+9}b_{k}\left( \nu \right) y^{\nu }$, with $%
b_{k}(0)=\gamma >0$ [18]. Let $\nu _{k}(1)$ denote the smallest integer for
which $b_{k}\left( \nu _{k}(1)\right) \geq 0$ and $b_{k}\left( \nu
_{k}(1)+1\right) <0$. Then we obtain the numbers $\nu _{k}(1)$ and $\nu
_{k}(2)$. These numbers satisfy the relations [18]
\begin{equation}
\frac{\nu _{k}(s)}{k-\nu _{k}(s)+1}\leq \xi \left( k,\nu ,s\right) \leq 
\frac{\nu _{k}(s)+1}{k-\nu _{k}(s)},s=1,2,
\end{equation}
\begin{equation}
\lim_{k\rightarrow \infty }\frac{\nu _{k}(s)}{k-\nu _{k}(s)+1}%
=\lim_{k\rightarrow \infty }\xi \left( k,\nu ,s\right) =y_{s}>0,s=1,2,
\end{equation}
\begin{equation}
f\left( y_{s}\right) =0,s=1,2.
\end{equation}

Therefore it is always possible to construct a convergent sequence to obtain
the positive roots of Eq.(\ref{26}). In the case of the neutral quark fluid, $q(u)\equiv 0$ and Eq. (\ref{26}) is reduced to a sixth order algebraic equation also having two
positive roots.

\section{Conclusions}

In the present paper we considered the collapse of a strange quark fluid in
Vaidya geometry. The possible occurence of a central naked singularity has
also been investigated and it has been shown that, at least for a particular
choice of the parameters, a naked singularity is formed. Depending on the
initial distribution of density and velocity and on the constitutive nature
of the collapsing matter either a black hole or a naked singularity is
formed. The values of the parameters in the solution (\ref{18})-(\ref{21}) determine which of these possibilities occurs. The solution
describing the collapse of the quark matter is not asymptotically flat and
this condition does not play any role in the formation of the naked
singularity. Due to the presence of the bag constant $B$ (playing the role
of a cosmological constant, from a formal mathematical point of view) the
mass function (\ref{18}) gives a cosmological type metric.

Quark matter, deconfined phase of hadronic matter at high temperatures or
densities may reside as a permanent component of neutron stars core or to
form stable compact stellar objects [2,4,5]. In fact from a physical point
of view it seems to be one of the best and more realistic candidates for the
study of properties of collapsing objects. It also serves to ilustrate the
much richer interplay that can occur among particle physics and general
relativity when more involved quantum field theoretical models are
considered.

\section*{Acknowledgments}

The authors are very grateful to the referee for many helpful comments,
which helped to improve the manuscript. This paper is supported by a RGC
grant of Hong Kong government.

\end{document}